\documentclass{llncs}
\usepackage{amsmath}
\usepackage{amssymb}
\usepackage{amsfonts}
\usepackage{pifont}
\usepackage{bm}
\usepackage[all]{xy}
\usepackage{color}
\usepackage{multicol}
\usepackage[pdftex]{graphicx}
\usepackage{fullpage}
\usepackage{wrapfig}
\renewcommand{\bibitem}{\vskip 2pt\par\hangindent\parindent\hskip-\parindent}

\newcommand{\nin}{\noindent}

\newcommand{\beq}{\begin{equation}}
\newcommand{\eeq}{\end{equation}}
\newcommand{\bea}{\begin{eqnarray}}
\newcommand{\eea}{\end{eqnarray}}
\newcommand{\be}{\begin{enumerate}}
\newcommand{\ee}{\end{enumerate}}
\newcommand{\bi}{\begin{itemize}}
\newcommand{\ei}{\end{itemize}}
\newcommand{\bverb}{\begin{verbatim}}
\newcommand{\everb}{\end{verbatim}}
\newcommand{\bv}{\begin{verbatim}}
\newcommand{\ev}{\end{verbatim}}

\newcommand{\bl}{\begin{large}}
\newcommand{\el}{\end{large}}
\newcommand{\bL}{\begin{Large}}
\newcommand{\eL}{\end{Large}}

\newcommand{\I}{\mathbf{I}}

\newcommand{\A}{\mathbf{A}}

\newcommand{\ket}[1]{\left| #1 \right>} 

\begin{document}

\title{Some Quantum-Like Features of Mass Politics \\
in Two-Party Systems \\ {\small April 2, 2011} \\}
\author{Christopher Zorn\inst{1} \and Charles E.\ Smith, Jr. \inst{2}}
\institute{Department of Political Science, Pennsylvania State University, University Park, PA 16802, USA \and Department of Political Science, University of Mississippi, Oxford, MS 38677, USA}

\maketitle

\begin{abstract}
The rapid expansion of the quantum interaction (QI) community's research on cognitive processes, games, and models of social measurement is affecting every field in the social sciences, and political science is no exception.  This paper aims to expand the substantive terrain of QI's reach in the discipline by illuminating a body of political theory that to date has been elaborated in strictly classical language and formalisms but has complex features that seem to merit generalizations of the problem outside the confines of classicality.  The line of research, initiated by Fiorina in the 1980s, is of special interest to scholars seeking to understand the origins and nature of party governance in two-party political systems wherein voters cast partisan ballots in two contests, one that determines partisan control of the executive branch and another that determines party control of a legislature.   We describe how research in this area evolved in the last two decades in directions that bring it now to the point where further elaboration and study seem natural in the more general formalistic and philosophical environments embraced in QI research.  In the process, we find evidence using data from a 1990s-era study that a restriction of a classical model that has animated work in the field appears violated in a form that gives way naturally to embrace of the superposition principle.  We then try to connect classical distinctions between separable and nonseparable preferences that are common in political science to their quantum and quantum-like counterparts in the QI literature, finding special affinity for a recently-introduced understanding of the distinction that provides a passageway into the boundary between fully quantum and fully classical views of the distinction and thereby provides new leverage on existing work and data germane to the theory .  We conclude with a brief outline of a primary data collection effort presently underway that is animated by the measurement and modeling insights developed in this paper.
\end{abstract}

%

\section{Introduction}

\hspace{6 mm} Among all of the academic specialties customarily identified as social sciences, political science is perhaps the greatest ``debtor" discipline, in the sense that so many of the theories and methods and models put to the task of understanding politics are borrowed from scholars working in other fields.  It is thus predictable that some of the latest and most promising theoretical and methodological innovations providing insight into the operation of politics are not native to political science.  What is surprising is their footing in quantum mechanics.  Long thought in the main to be a theory with applications exclusive to the realm of the near-unobservably small, where probabilities rather than observable mechanics propagate in accordance with causal laws, the 21$^{\textrm{st}}$ century is becoming witness to an ever-growing export market for the quantum formalisms and the probability theory native to them.

This paper follows that trend by illuminating a body of political theory that to date has been elaborated in strictly classical language and formalisms but has complex features that seem to merit generalizations of the problem outside the confines of classicality.  The line of research, initiated by Fiorina in the late 1980s, is of special interest to scholars seeking to understand the origins and nature of party governance in two-party political systems wherein voters cast partisan ballots in two contests, one that determines partisan control of the executive branch and another that determines party control of a legislature.   We describe how research in this area evolved in the 1990s and 2000s in directions that brings it now to the point where further elaboration and study seem natural in the more general formalistic and philosophical environments embraced in QI research.  In the process, we find evidence using data from a 1990s-era study that a restriction of a classical model that has animated work in the field appears violated in a form that gives way naturally to embrace of the superposition principle.  We then try to connect classical distinctions between separable and nonseparable preferences that are common in political science to their quantum and quantum-like counterparts in the QI literature, finding special affinity for a recently-introduced understanding of the distinction that provides a passageway into the boundary between fully quantum and fully classical views of the distinction and thereby provides new leverage on existing work and data germane to the theory .  We conclude with a brief outline of a primary data collection effort presently underway that is animated by the measurement and modeling insights developed in this paper.

\section{Balancing Theory}

\hspace{6 mm} Here we address a body of theory that is of special interest to scholars seeking to understand the origins of party governance in political systems where voters cast partisan ballots in two contests, one involving the executive branch and one involving the legislature.  In the U.S., one of the most prominent strands of research in this area was initiated by Fiorina in the late 1980s and early 1990s (e.g., Fiorina 1996).  In contrast to classical, Downsian (1957) models, where voters with policy preferences that are more moderate than the positions staked out by parties in two-party systems choose (if possible) the closest of the two alternatives, Fiorina's thesis emphasizes the importance of the two institutional choices in U.S.\ politics -- the congress and the presidency.  In his model, voter desires for moderation can be realized by splitting the ticket -- voting Republican in one institutional choice setting and Democratic in the other.  Likewise, voters with more extreme positions can maximize their returns by choosing one party across both institutional contests.  This strand of research thus contrasts with the binary choice (Democrat (\emph{D}) versus Republican (\emph{R})) tradition from Downs by framing the problem as a choice set for party governance ($G$) across four mutually exclusive options, $G = [D_E D_L, D_E R_L, R_E D_L, R_E R_L]$, where the subscripts distinguish the election contesting control of the executive branch from the one deciding control of the legislature.

Fiorina's initial formulation of the problem defined the choice options as well as voter positions relative to them in a one-dimensional, policy-specific, liberal-versus-conservative Euclidean space.\footnote{A generalization of that model to $N$ dimensions is straightforward, but specifications of its empirical implications relative to the four partisan choice options are not easily defined in a parsimonious fashion.}  Across indivduals, different issues have different (or no) salience.  Moreover, individual understandings/predictions of where the parties stand on issues may be variable.  For one or all of these reasons, measured policy preferences in the mass public are not stable across time, an empirical regularity traceable back at least to Converse (1964).  Another complication is that, a priori, the universe of salient policies in an election is difficult to determine, and thus measure, for all voters/respondents/subjects.  Given all these givens, it is perhaps not surprising that many of the scholars who have investigated the empirical relevance of Fiorina's ``policy balancing'' theory report that it provides little or no observable, explanatory purchase to our understanding of partisan or bipartisan (i.e. ticket-splitting) choice (cf.\ Petrocik and Doherty (1996), Carsey and Layman 2004).

However, ``party balancing" is a different matter.  As explained by Smith et al.\ (1999), ``the act of `policy balancing' implies that individual voters ultimately engage in `party balancing'" (p.\ 742), a process whereby voters adjust their preferences regarding which party should control one institution based on either preferences for or expected outcomes about partisan control of the other.  The focus of this study was narrow: the authors took as their primary task an analysis of how then-customary, statistical models of candidate/party choice in U.S.\ congressional elections might be better specified given an account of measurement metric implications derivative of one (of several) possible, theoretical exposition(s) of party balancing.  However, both the theory underlying the hypotheses tested in the research and the data used to do so are perhaps of broader interest.  On the theoretical side, this study leans on one account of how social scientists understand the distinction between preference separability and nonseparability, issues that merit attention given their kinship (and lack thereof) with the quantum mechanical meanings of those terms. On the empirical side, results in the survey data used in the Smith et al.\ study are not readily accommodated by classical formalizations.  In the subsections that follow, we elaborate.

\subsection{Classical Views of Separability and Nonseparability}

\hspace{6 mm}Social scientists understand and use the words separable and nonseparable in ways that are distinct from the quantum mechanical meanings of the terms.  The most general tradition uses the terms to distinguish between two types of preference orders.  The most basic, classical example is one voter with two considerations, observable as two bits.  Preferences are said to be separable when each of those preferences arises independent of the consideration of or outcome on the other.  Of the twenty-four (4!) possible preference orderings in the two-bit example, the eight orderings with last preferences as mirror images of the first (e.g.\ 00 ... 11) are understood to be separable orderings when the considerations are of equal salience and the orderings are observed across groups of voters as invariant to the order in which the preferences are measured (Lacy and Niou 1998).

\begin{wrapfigure}{l}{75mm}
 \centering
 \caption{Divided vs.\ Unified Control of Government} \label{ExecLeg}
 \includegraphics[width=75mm]{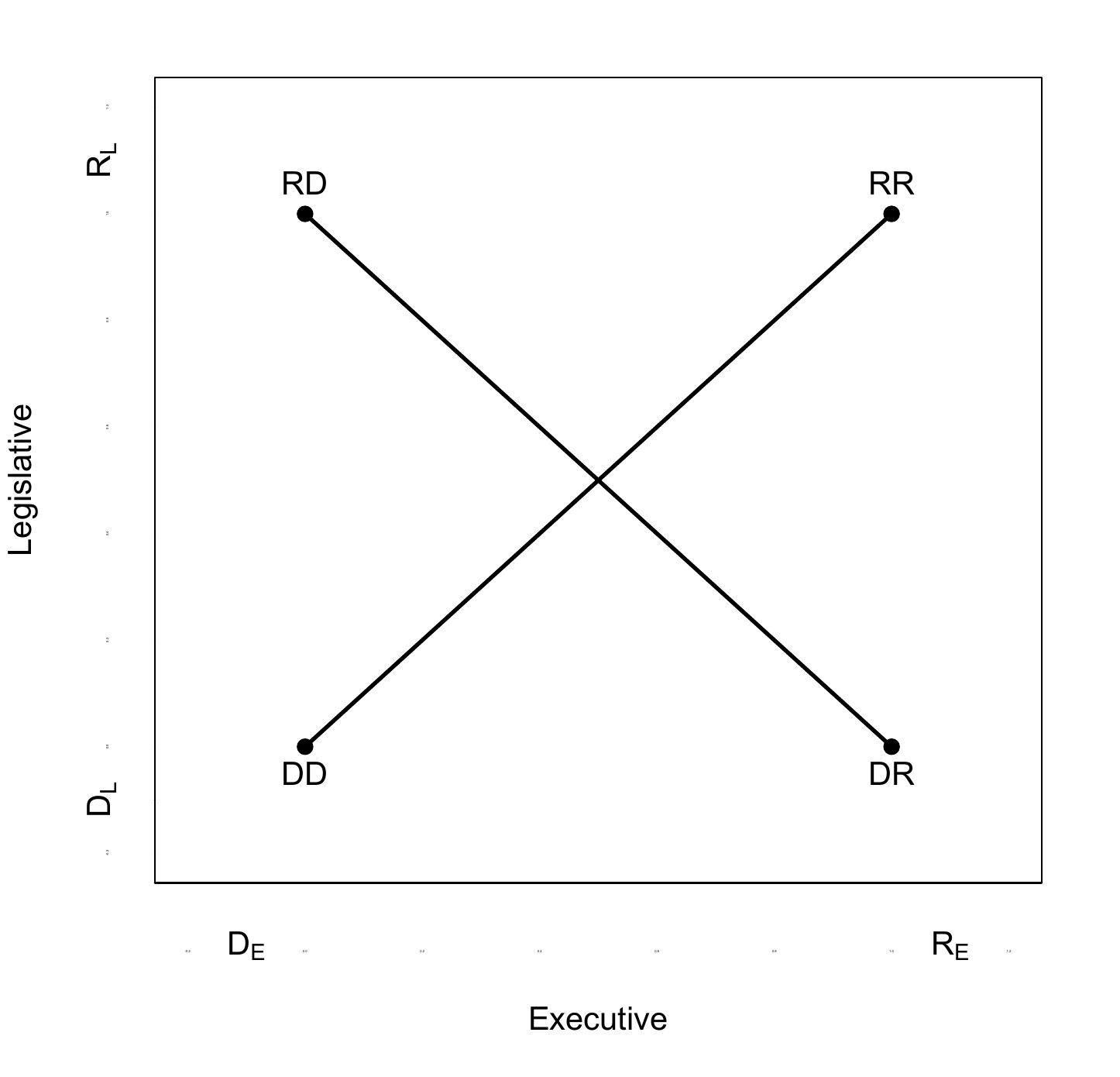}
\end{wrapfigure}

A visually intuitive and simple alternative to understanding separability and nonseparability in previous social science work (including that on balancing theory) is animated by a simple model and illustration.  For Figure \ref{ExecLeg}, define $S$ as an initial state belief vector that can be used to describe considerations over preferences regarding partisan control (Republican versus Democratic) of both the executive and the legislative branches in an election.  Belief vectors regarding partisan options in the two-dimensional space can then be described in terms of coordinates specific to each branch.  Further, define a simple Euclidean distance in the space:

\beq \label{EucDist}
||S_{E} - S_{L}||_{\I} =\sqrt{(R_{E}-D_{E})^2+(R_{L}-D_{L})^2}
\eeq

\noindent where \textbf{I} is $2 \times 2$ with elements that can be interpreted as weights.  Specifically, the main diagonal weights signify the salience of the two, associated dimensions of party governance, their ratio the relative importance of them.  In order for the space to remain Euclidean, the off diagonal elements must be equal.  When as here they are jointly equal to zero, ``there is no interaction between'' (Enelow and Hinich 1984, 19) the considerations, and the preferences arising from them are said classically to be separable.

Now consider an alternative transition matrix:

\bea
\A &=&
\left[{\begin{array}{cc}
 \alpha_{11} \hspace{3mm}
 \alpha_{12} \\ 
 \alpha_{21}\hspace{3mm}
 \alpha_{22}
 \end{array}}\right]
\eea

\noindent with the restriction ${\alpha_{12}} = {\alpha_{21}} = {\alpha}$.  Replacing \textbf{I} with \textbf{A} gives:

\beq \label{WtdEucDist}
||S_{E} - S_{L}||_{\A} =\sqrt{{\alpha_{11}}(R_{E}-D_{E})^2+2{\alpha}(R_{E}-D_{E})(R_{L}-D_{L})+{\alpha_{22}}(R_{L}-D_{L})^2}
\eeq

\noindent This is a stylized version of the weighted Euclidean norm developed by Enelow and Hinich (1984).\footnote{Hinich and Munger (1997) generalize the norm to $N$ dimensions.  Spatial voting theory more generally is built upon the early work of Davis and Hinich (e.g., 1968); Gorman (1968) is a fundamental work on the concept of separability, an idea he credits without specific citation to Leontief.  Schwartz (1977) was first to consider the problem in multiple elections.  Lacy (2001a,b; Lacy and Niou 1998) offers more recent examples of applied and theoretical work on the separability-nonseparability distinction in political decision theory.}  Fixing coordinates at the poles of one dimension and differentiating the square to invoke preferences in the other when $\A$ is a matrix of positive ones gives $R_E D_L$ and $D_E R_L$.   These are the choice options of the balancer -- the voter who prefers a form of coalition government to one-party control.  Conversely, preferences in the unifying regime -- $R_E R_L$ and $D_E D_L$ -- are given by fixing the off-diagonal elements of $\A$ at -1.  Classically, these conditions imply that preferences are fully nonseparable and assume that the relevance to party governance of the executive and legislature are equal.  The probabilities of the four outcomes for partisan control of government thus sum to unity when given $\pm\alpha$ and a partisan outcome in one dimension.  That is, for a $4 \times 1$ state vector ($\psi$),

\bea \label{psiNegAlpha}
\psi &=&  
\left[ {\begin{array}{cc}
 \psi DD  \\
 \psi DR  \\
 \psi RD  \\
 \psi RR  \\
 \end{array} } \right]
 \longrightarrow
\hspace{6mm}\psi_{-\alpha} = \frac{1}{2} 
\left[ {\begin{array}{cc}
 1  \\
 0  \\
 0  \\
 1  \\
 \end{array} } \right],
\eea

\nin the vector transitions for voters with $-\alpha$ to the unifying regime and $\Pr( DD \mid D ) = \Pr( RR \mid R) = .5$.  Likewise, $\psi$ transitions to the balancing regime and 
$\Pr( RD \mid D) = \Pr( DR \mid R) = .5$ with positive $\alpha$, as in:

\bea \label{psiAlpha}
\psi =  
\left[ {\begin{array}{cc}
 \psi DD  \\
 \psi DR  \\
 \psi RD  \\
 \psi RR  \\
 \end{array} } \right]
 \longrightarrow
\hspace{6 mm}\psi_{\alpha} = \frac{1}{2} 
\left[ {\begin{array}{cc}
 0  \\
 1  \\
 1  \\
 0  \\
 \end{array} } \right].
\eea

Most modeling and analysis in the social sciences proceed implicitly from the assumption that the observables involved in the models arise from separable considerations even when there are sound theoretical reasons to expect otherwise: economists aggregate goods in bundles that might not be separable in the minds of consumers, while political scientists do the same thing with issue preferences and voters.  At best, these practices foreclose nuanced observation of potentially interesting phenomena; in the worst scenarios, they may lead to faulty inferences.  As a result, it is easy to embrace theoretical and empirical work in the social sciences that make and/or test explicit assumptions about the distinction between separability and nonseparability.  We are however given pause with respect to certain aspects of modeling and nomenclature conventions used routinely in social science in the course of defining and understanding the distinction.  In the subsection that follows, we show evidence using the Smith et. al data that a tenet of a classical model that has animated work in the field appears violated in a form that gives way naturally to embrace of the superposition principle and then suggest that the classical formalisms and theories of preference separability might best be viewed as special cases of the quantum versions even when the real-world problems being studied are decidedly macroscopic and traditionally seen in the social sciences only through the lens of classicality.

\subsection{Toward a More General Framework}

\hspace{6 mm}Spatial representations of voting have been criticized for being overly restrictive, not least with respect to defining separable and nonseparable preferences.  Lacy (2001), for example, notes that in $N$ dimensions, the model set out by Hinich and Munger (1997) does not account for the possibility that sets of preferences might be nonseparable from one or more other sets.  He also implies that the symmetry of the transition matrix $\A$requires, given nonseparability, that each preference depend equally on the outcome relative to the other consideration.  We have come to see these sorts of criticisms as wide of the mark.  The first critique can be addressed simply by further generalizing the Hinich-Munger model.  The second is only true if, as in our example above, it is assumed to be so; the weighted Euclidean norm certainly does not require, to reference our example, that the legislative preference be conditioned by the executive outcome when the reverse is true (formally: one of the main diagonal weights in $\A$ can be zero while the other is nonzero).  Indeed, where others see differences in generality across the spatial and preference-order representations of separability and nonseparability in social science work, we, excepting presentation differences, see commonality.  One tradition, the spatial theory, lays bare via a toy model the mechanics of the distinction; the other eschews a continuous, interval level metric; in two dimensions such as the problem above, they both distinguish in question order experiments or conditioning questions the same preference orders as exhibits of separability, the same orders as exhibiting nonseparability.  Thus, at least relative to the balancing problem in two dimensions, the social science traditions are equivalent.

\begin{wraptable}{r}{80mm}
\caption{Conditional and Unconditional Preferences for Congressional Control, 1996} \label{crosstabs}
\begin{center}
\begin{tabular}{lcccc}
\hline \hline
Unconditional  \hspace*{8pt} &   \multicolumn{3}{c}{\underline{Democratic Executive}} &  \\
Preference            &   Dem.\  &   Either & GOP & Total \\
\hline
Democrat            &   188 & 16 & 98  & 302  \\
Either/DK             &   77  &  26 & 151 & 254  \\
GOP                    &     33  &  9  &  339 & 381  \\
                           &          &       &         &         \\
Total                   &   298 & 51  &  588 & 937  \\
\hline \hline
\multicolumn{5}{c}{$\chi^{2}_{4} = 256$ $(P<0.001)$; $\hat{\gamma} = 0.72$.} \\
 & & & & \\
 & & & & \\
\hline \hline
Unconditional        &   \multicolumn{3}{c}{\underline{GOP Executive}} &  \\
Preference            &   Dem.\  &   Either & GOP & Total \\
\hline
Democrat            &   258 & 10 & 34  & 302  \\
Either/DK             &   151  &  31 & 74 & 256  \\
GOP                    &     99  &  14  &  266 & 379  \\
                           &          &       &         &         \\
Total                   &   508 & 55  &  374 & 937  \\
\hline \hline
\multicolumn{5}{c}{$\chi^{2}_{4} = 290$ $(P<0.001)$; $\hat{\gamma} = 0.73$.} \\
\end{tabular}
\end{center}
\end{wraptable}

What the traditions in social science have most in common is their firm and exclusive footings in classicality.  This is most readily illustrated in the context of the balancing problem with a simple hypothetical.  Assume survey respondents are asked two conditioning questions about their preferences regarding partisan control of each of the two institutions, and further that all respondents choose $R$ across all four conditions.  These observables would reveal one preference order for all subjects, with $RR$ most preferred and $DD$ least.  As such, they would satisfy the inverse rule (Lacy 2001a,b), which in two dimensions is a necessary and sufficient condition to establish separability via the preference order tradition of understanding these terms.  From the standpoint of the spatial theory of voting, the off-diagonal of \textbf{A} would be presumed revealed as zero, and likewise, the preferences called separable.  More fundamental, though, is that classically-trained social scientists would assume that something else is revealed in these observations, namely the probabilities that the respondents would have preferred $R$ for each of the institutions in the absence of the conditioning questions.  Axiomatically, the probability of $R$ absent the conditions is a weighted average of the conditional probabilities gauged across the mutually exclusive and exhaustive options of the conditions.  In our hypothetical, the unconditional probability of $R$ as a revealed preference must then be unity in both dimensions.

A close reexamination of the data used by Smith and his colleagues in their 1990s study of the balancing problem gives pause against full embrace of such an axiom.\footnote{The data for the Smith et al.\ study came ``from a pre-election telephone survey conducted by the Social Science Research Laboratory at the University of Mississippi between October 11 and November 3, 1996.  The sample covered the lower forty-eight [U.S.] states and the District of Columbia.  The data set contains 995 completed observations.'' (Smith et al.\ 1999, p.\ 739)}  Following their Table 2 (p.\ 748), we report results in a contingency table from two conditional questions about partisan control of the U.S.\ Congress, one fixing a Republican victory in the presidential election and one fixing a Democratic win; both are compared to a variable measuring each respondent's ``unconditional'' preference over partisan control of Congress (that is, without conditioning on executive control).  Results on the main diagonal of the tables thus denote respondents who answered consistently across the two conditions, and are described as characterizing voters with separable preferences.  This is true by any classical standard, but we wondered about the purchase of the likewise classical assumption about preferences absent the conditions.  The study includes and features two indicators proposed as such, one fashioned by the authors and called a ``direct" measure, and another, ``indirect" indicator that is among the most familiar measures in U.S. political science, the party identification measure developed by the authors of \textit{The American Voter} (Campbell et al.\ 1960) and used continuously since in the biannual U.S.\ National Election Studies.  We look now at both unconditional preference measures within the subset of respondents classified as having separable preferences based on their consistent partisan choices across the conditions and find provocative results.

Judged against the unconditional legislative preference measure developed by the authors, almost 16\% of all survey respondents in the study are classified as having separable preferences using the conditional measures but offer a different preference absent the conditions.  As a percentage of those classified as having separable preferences, respondents with different unconditional preferences count north of 20\%.  A large measure of the effect is owed to the authors' inclusion of middle categories in the preference measures, and to the respondents' choosing $DD$ or $RR$ given the conditioning but a neutral position in the absence of it.  However, there is a nontrivial amount of outright party switching (conditional to unconditional) in these data:  8.8\% of those conditioned to $DD$ chose $R$ absent the conditioning; 4.7\% conditioned to $RR$ choose $D$.  Using party identification as the unconditional measure, nearly thirty percent of voters classified as having separable preferences give a response different from the consistent ones they give on the conditional indicators.  As with the other measurement standard, the partisanship version of unconditional party preference shows that the prevalent quirk in the data is the tendency for $DD$ voters to chose a more Republican option without the conditioning.  Indeed, the proportions of $DD$ voters who seem more Republican in the unconditional measures are statistically distinct from the proportions of $RR$ voters who seem unconditionally more Democratic ($P=.011$ using the ``direct" measure of unconditional preference, $P=.016$ using party identification).  If this is measurement error, it does not appear to be random, as we would expect these proportions to be indistinguishable from each other given the symmetry of the problem's context.

Obviously, these results do not comport well with the classical understandings of separability.  Classical rationales for the misfit data are certainly possible; measurement error, for example, need not be random.  However, if we ponder such extra-contextual explanations while maintaining the classical theoretical motivation for the problem, it is important to recognize that potentially important substantive insights inside the context of the problem are foreclosed.  One we have in mind is rooted deeply in the normative concerns presumed to motivate voters to on the one hand divide or on the other unify partisan control of the branches.  Another relates to the nature of partisanship itself, originally and still-usually construed as an affective orientation toward a group.  With respect to the distinction between favoring one-party versus divided control, the fact is that there are plausible reasons to favor both, plausible reasons even to feel disfavor toward both.  In the U.S. system, one-party control is described pejoratively by some as undermining the system of checks and balances designed to limit the power of government, or at least the speed with which it affects sweeping change.  One-party control is described elsewhere as necessary to avoid gridlock, and better still because in its absence voters have difficulty behaving responsibly in rewarding positive performance and punishing failure.  Divided control can likewise sensibly be described as good or bad, or both at the same time.  And so it is as well with the parties.  Positive affect toward one party need not be accompanied by an equal amount of antipathy toward the other.  Judged anecdotally from public discourse, antipathy toward both parties in the U.S.\ is widespread, though of course we can also imagine voters with positive affect toward both of them at the same time.

The clear message is that this problem is as simple as elementary geometry when applied to the institutions and the parties, but potentially very complex in the minds of voters.  At a given time, partisan control of the U.S.\ government (and like systems) will in fact either be divided across the parties or unified by the leadership of just one.  Likewise, a given branch of the U.S.\ government (and other two-party systems) will at a given time be controlled either by one party or the other.  These realities yield exhaustive and mutually exclusive conditions, and classical models that superimpose the conditions that follow from these externalities upon voters have an attraction -- a verisimilitude judged against political arrangements fixed in specific ways.  But, as we have seen, they are limited.  Indeed, in the context of the balancing problem, it is awkward to write down preference orders and, a priori, single some out as evidence of separable preferences and others as markers of nonseparability if the cognitive processes at work in the minds of voters as they reveal their preferences might be more nuanced than the rationale for labeling the orders as separable or not.  Similarly, the spatial model provides no quarter, as it will, for example, accommodate an induced voter preference for $R_{L}$ from $D_{L}$ in one regime, but not in the other -- this no matter how deep our suspicions that some voter might maintain affect or antipathy for both regimes at the same time.  

Fashioning variables from the answers to a pair of agree-disagree items that were part of the survey instrument used in the Smith et al.\ study but never reported on provides further insight.  Specifically, the instrument queried respondents about their affect toward unified and divided government in separate items, one gauging agreement/disagreement with the statement ``the government works best when one party controls both the Congress and the presidency,'' and another with the opposite stimulus, positing that ``the government works best when control of the congress and the presidency is split between the parties.''   Per restrictions in the theory, we would expect that variables fashioned to order increasing agreement with these statements would, for a voter with nonseparable preferences, elicit high agreement on one accompanied by low agreement on the other, and that across voters, these variables would be highly and negatively correlated.  We find instead that that they are not related in a statistically discernible way, and this is true both in the main, and within groupings of respondents that exhibit on the one hand alignment in their partisan preferences for control of the Congress given partisan outcomes for the presidency and on the other among respondents that chose different parties for control of the legislature across the conditioning executive options.  Across individuals, indeed, affect toward unified control appears quite orthogonal to affect toward divided control\footnote{The Pearson's correlation between the two variables is -0.05 ($\hat{\gamma} = -0.07$).} -- a result far afield from the theory.

\begin{wraptable}{r}{75mm}
\caption{Influences on Separability of Preferences} \label{logits}
\begin{center}
\begin{tabular}{lcc}
\hline \hline
            &   Two-Item &   Three-Item \\
            &   Response  &   Response   \\
\hline
Constant    &       -0.47&        0.53\\
            &      (0.28)&      (0.29)\\
Affinity: Unified &       -0.32&       -0.35\\
            &      (0.10)&      (0.11)\\
Affinity: Divided &        0.56&        0.54\\
            &      (0.11)&      (0.11)\\
Democrat    &       -0.27&       -0.54\\
            &      (0.17)&      (0.19)\\
Republican         &       -1.14&       -1.42\\
            &      (0.18)&      (0.19)\\
            &               &               \\
$\chi^{2}_{4}$  & 91.8 & 113.4 \\
$N$             & 899      &  847    \\
\hline \hline
\end{tabular}
\end{center}
\end{wraptable}

Still more insight into the fit of the classical assumptions to the data is given by fashioning binary variables from the legislative preference measures that would seem to distinguish respondents who exhibit preferences consistent with classically separable preferences from those exhibiting classical nonseparability.   We did this two ways.  One -- identified as the ``two-item response" variable in Table \ref{logits} -- is scored zero when both executive conditioning questions evoke identical legislative preference reports across the conditions and one otherwise.  The second (``three-item response'') requires in the zero category the same as the first, plus a third, unconditioned match to the two identical reports given the conditions.  What these variables do is separate respondents whose legislative preferences appear invariant to conditioning with hypothetical executive outcomes (and thus presumed separable from \emph{E}) from those who exhibit adjustments (perhaps nonseparable from \emph{E}).  As an additional pair of controls, we also include indicators for respondents' partisan self-identification; \emph{Democrat} and \emph{Republican} are naturally-coded indicator variables gleaned from an ANES-like measure of party identification. Inclusion of these variables obviates the potential that our results for unified / divided government are driven by purely partisan concerns (as would be the case, for example, if Democrats -- whose incumbent president held a clear lead over his Republican challenger for most of the 1996 election season -- were more solicitous of divided government than were Republicans). 

We find in Table \ref{logits} that both affinity for \emph{Divided} government and affinity for \emph{Unified} control -- variables uncorrelated with each other -- contribute in statistically relevant ways to these distinctions, with proponents of unified control having a greater likelihood of also exhibiting separable preferences; in contrast, those with greater affect toward divided control of government were more likely to reveal legislative preferences that seem conditioned by (and thus nonseparable from) outcomes on \emph{E}.  These results, as with the apparently orthogonality of the affinity measures, are not animated by the classical theory.  Likewise, turning again to Table \ref{logits}, we find puzzling results with partisanship amidst the covariates.  The main result is that self-identified Republicans seem especially invariant to the conditioning items.  Here again, we see no accommodation for such an effect in the formal theory that animates the Smith et al.\ study (or any other study), and thus conclude that the balancing problem is ripe for generalization.

\section{Testing a Pseudo-Classical Model of Voter Preference}

\hspace{6 mm}Our task, then, is to develop a model of partisan balancing that can reconcile existing theoretical accounts with empirical reality.  In doing so, however, we do not wish to fully foreclose on the simple model of \S 2.1 and its implications, first because it is clear from the data we and others have examined that this model ably characterizes the preferences of many voters, and that its distinctions between unifying and dividing have important empirical implications for statistical models of voter choice.  Second, because an on-going interest in our research program is to better understand and articulate translations of models that have to-date been viewed in political science from strictly classical perspectives, we wish here to keep the one in \S 2.1 prominent in the background for the purpose of illuminating, if here in only a preliminary way, what kinships we can divine between the classical treatment of the problem and prominent advances in the QI literature.  We begin by noting that there is no dispute in the U.S.-based political science literature over the dimensionality of the balancing problem, or others like it; indeed, in every paper we have seen that broaches the topic of separable versus nonseparable preferences, a simple example of what can be called the two-bit case is referenced in the course of explaining the distinction.  However, as Smith et. al note, the weighted Euclidean has implications for the metrics of the space.  The authors trace one implication into an analysis of the fit of statistical models, but as with others writing before and after them, they otherwise treat the weighted Euclidean as a tool to classify voters against hypothetical arrangements of parties and institutions -- these voters evidencing separable preferences, these showing nonseparable preferences of a particular form, and so on.  They do not, and indeed no classical scientist has, considered it as a cognitive process model, descriptive of the thinking of an individual voter.

When we do so, we see an important change in the dimensionality of the space when the elements of $\bf{A}$ are nonzero.  Indeed, when as in \S 2.1 $\bf{A}$ is a matrix of ones, we see the problem applied to the voter as producing four and not two bits.  This is because there are at the same time for every voter two partisan preferences for each dimension of governance, one (say, for the $L$egislature) ``invoked" by a fixed (say, $R$) outcome in the other ($E$xecutive), and a second one, also for the legislature, ``invoked" by the opposite fixed outcome ($D$ = $E$xecutive).  So, for every voter, $L$ can be conceptualized as a two-bit registry, and likewise for $E$, making the total four.  Conceptually, we are now only a step away from a full alignment of the problem in a more general space, and indeed are outside the bounds of classical approaches already in considering two partisan preferences for each dimension at the same time for one voter.  From a quantum perspective, such preferences would be described as being in superposition, and in a Hilbert space, the problem we have elaborated here would be described not in terms of four bits but rather in terms of two qubits.  Such a space would generally have dimension $2^{k}$, where \emph{k} is the number of qubits.  Defined over the real numbers, this space has deep kinships with spatial models in political science, including its depiction of distance, which is Euclidean.  Defined over complex numbers, the problem is made still more general, as we would then have voters consider executive ($e$) and legislative ($l$) dimensions to governance and define partisan options ($r$ and $d$) for each, writing the tensor product:

\beq \label{KETS}
\ket{e} \otimes \ket{l} = a_0 b_0 \ket{dd} + a_1 b_0 \ket{dr} + a_0 b_1 \ket{rd} + a_1 b_1 \ket{rr}
\eeq

\noindent where the two qubits are in superposition, $e$ and $l$ are independent, and the existence of the full complement of product weights defines separability.  

In \S 2.1 we wrote of differentiating the square of the weighted Euclidean and finding four partial derivatives to invoke completely nonseparable preferences, one set shifting voters to the balancing regime and the other shifting other voters to the unifying regime.  The quantum analogy of the four outcomes is:

\beq
\nonumber \frac{1}{\sqrt{2}}(\ket{dd}+\ket{rr})\hspace{10 mm} \frac{1}{\sqrt{2}}(\ket{dd}-\ket{rr})
\eeq

\beq
\nonumber \frac{1}{\sqrt{2}}(\ket{dr}+\ket{rd})\hspace{10 mm}\frac{1}{\sqrt{2}}(\ket{dr}-\ket{rd})
\eeq

\noindent These are the maximally-entangled, two-qubit states namesaked for John Bell after his fundamental work (1964) on the Einstein-Podolsky-Rosen paradox (1935).  In the QI literature, this type of nonseparability has been considered in application to social/cognitive data for more than a decade (e.g., Aerts et al. 2000;  Aerts et al. 2005; Bruza et al. 2009; Bruza et al. 2010;  Gabora and Aerts 2009), and is of a radically different nature than any corollary ever considered in political science.  Formally, ``there are no coefficients which can decompose'' (Bruza et al.\ 2009, p.\ 6) the states into the tensor product above that sets out $e$ and $l$ as independent.  When conceptualized as resident in a Bell state, quantum nonseparable, preferences in one dimension are not so much conditioned or dependent upon outcomes in another as given by them, so even to properly consider those in one dimension requires consideration of those in the other.  

Likewise, quantum separability differs radically from separability conditions customarily considered in political science.  As fashioned above, the tensor product $\ket{e} \otimes \ket{l}$ seems to capture quite precisely the language used by political scientists when describing separability, as considerations in one dimension are independent of those in the other at the level of the voter.  However, we have come to recognize that separable preferences as defined in the political science literature would not necessarily be viewed as separable in the quantum generalization.  This can be seen in the tradition of preference order rankings understanding of separability by noting that one could not, against the quantum definition of separable preferences, write out a subset of the 4! preference orders in the problem and, a priori, privilege eight or four or even one of them as demonstrating separability.  Likewise, classical reasoning from the spatial model and the weighted Euclidean about separability runs aground against the mathematics in \eqref{KETS}.  Indeed, the weighted norm fashioned so as to depict fully separable preferences they are traditionally understood in political science can be readily interpreted as mapping directly to one of the Bell states.  The traditions, quantum versus classical, thus seem at once deeply related, and profoundly incompatible.

However, a recent and we think quite important innovation by Bruza, Iqbal, and Kitto (2010) provides a passage into the boundary between Bell-type entanglement and nonseparability as it has been traditionally understood in political science.  Challenging what perhaps was a status quo in the QI literature -- using the Bell inequalities as ``the formal device for determining non-separability" (p. 26) -- the authors add to the nomenclature the notion of ``psuedo-classical" nonseparability and situate understanding of this phenomenon in territory familiar to classically-trained social scientists by formalizing it as a factorization problem in a joint probability distribution.  Probabilistically, if generalizing balancing theory fully quantum, we would refashion the state vector ($\psi$) from 2.1 to an uninformed state ($\psi_{u}$): 

\bea
\psi &=&  
\left[ {\begin{array}{cc}
 \psi DD  \\
 \psi DR  \\
 \psi RD  \\
 \psi RR  \\
 \end{array} } \right]
 \longrightarrow
\hspace{6 mm}\psi_{u} = \frac{1}{2} 
\left[ {\begin{array}{cc}
 1  \\
 1  \\
 1  \\
 1  \\
 \end{array} } \right]
\eea

\noindent where probabilistic reasoning must shift from within the confines of classical, Kolmogovorian theory to that of the more general theory often namesaked for Born (1926).

The Bruza et al.\ (2010) innovation in contrast foots the distinction between separable nonseparable in classical probability theory, retaining the law of total probability that is not a feature of the fully quantum perspective.  Elaborating from a theorem proved by Suppes and Zanotti (1981), Bruza et al.\ (2010) note that for two random variables $A$ and $B$ and a conditioning (factorizing) variable $\lambda$, 

\beq \label{CondAB}
\Pr(A,B,\lambda) = \Pr(A|\lambda) \Pr(B|\lambda) \Pr(\lambda)
\eeq

\nin and 

\beq \label{UncondAB}
\Pr(A,B) = \sum_{j \in \Lambda} \Pr(A|\lambda_{j}) \Pr(B|\lambda_{j}) \Pr(\lambda_{j})
\eeq

\nin where $\Lambda$ is the set of values taken on by $\lambda$.  In this framework we find a new lever into the balancing problem and the Smith et al. data by considering our executive party priming variable $\lambda$, and the respondents' choices over partisan Congressional control our central variable of interest $Y$.  In analogous fashion to \eqref{UncondAB}, we can write

\beq \label{UncondY}
\Pr(Y) = \sum_{j \in \Lambda} \Pr(Y|\lambda_{j}) \Pr(\lambda_{j}).
\eeq

\noindent By treating the marginals of our unconditional Congressional control measure as an empirical estimate of the ``true'' unconditional distribution, we can compare (via a standard chi-square test) the cell frequencies for the two conditional measures to that for the unconditional item.  

\begin{wrapfigure}{l}{75mm}
 \centering
 \caption{$\chi^{2}$ Values for Conditional vs.\ Unconditional Preferences Over Divided Government} \label{ChiSquarePlot}
 \includegraphics[width=75mm]{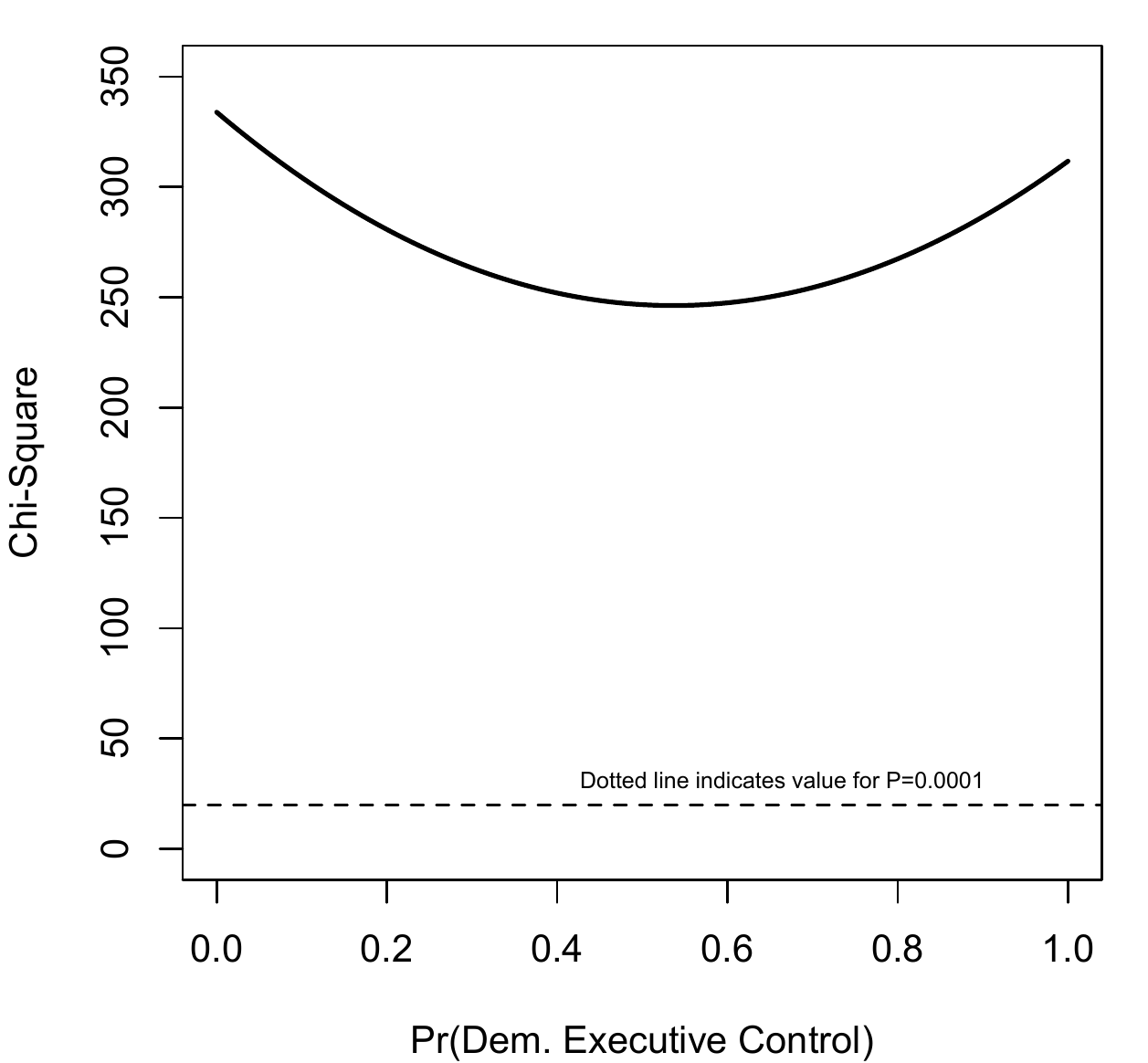}
\end{wrapfigure}

Bruza et al.\ note that, in addition to the law of total probability, their approach requires attention to the presumed prior probability distribution of $\lambda$, in particular that the distribution of $\lambda$ is uniform.  While in their experiments they randomly assigned subjects to priming conditions, here all respondents answer all three versions of the Congressional control question (conditional on Democratic control of the executive, conditional on Republican control of the executive, and unconditional).  As a result, to ensure the robustness of our findings we consider the range of possible values for the prior on $\lambda = \Pr(\text{Dem. Executive Control})$; consistency in the findings of the test across a broad range of potential prior values for $\lambda$ would suggest that our results are not sensitive to the choice of prior.

Figure \ref{ChiSquarePlot} plots the values of that $\chi^{2}$ statistic over a range of values $\tilde{\lambda} \in [0,1]$; cell frequencies for the statistic were thus calculated as $\tilde{\lambda} f_{Dj} + (1-\tilde{\lambda} f_{Rj})$, where $f_{Dj}$ and $f_{Rj}$ denote the cell frequencies from the one-way table of responses conditional on Democratic and Republican control of the executive, respectively. For all possible prior values of $\lambda$, we note a  substantial statistical difference between the distributions of preferences over partisan control of Congress between the conditional and unconditional measures, and at no point over the range of those values does the test statistic remotely approach statistical insignificance.  As Bruza et al.\ (2010, 2011) note, if the remaining two assumptions about the prior distribution of $\lambda$ and the law of total probability hold, this can be interpreted as evidence in support of nonseparability in preferences.

\section{Concluding Comments}

\hspace{6 mm} In a recent paper, Busemeyer et al.\ note that ``quantum information processing principles provide a viable and promising new way to 
understand human judgment and reasoning'' (2011, 54).  Somewhat more specifically, Bruza and colleagues suggest that their notion of ``psuedo-classical" nonseparability ``is a useful one in order to classify quantum-like systems" (Bruza et al. 2010). We are fully aligned with both of these these sentiments, and indeed have come to suspect that the latter will receive nontrivial acclaim as an outright alternative to what we have now come to view as very restrictive accounts in political science. 

At the same time, we recognize that we cannot advance our work on balancing much further in the absence of primary data.  We believe our reanalysis of the Smith et al.\ data is informative, but the fact that their data was not collected specifically for purposes put to task here means it has significant drawbacks.  Far and away the most obvious of these is that, while Smith et al.\ condition (and do not condition) their questions about Congressional control on partisan control of the executive branch, they neglect to do the reverse.  As a result, we cannot ascertain whether or not the same dynamics apply to executive control, and moreover, are limited (by precisely half) in our ability to model nonseparability in the full context of the problem.  Second, Smith et al.\ did not vary the order in which the three Congressional control questions (two conditional, one unconditional) were asked.  As many scholars have noted (e.g., Trueblood and Busemeyer 2011), a key implication of quantum-like macro-level phenomena are order effects; failure to account for such effects presents a serious risk to inference about such phenomena.  Last, some of the relevant measures in the Smith et al.\ study are fashioned so as to foreclose a superposition-like multidimensionality; party identification is but one example.

\begin{wrapfigure}{l}{75mm}
 \centering
 \caption{Current Divided Government Survey} \label{Qualtrix}
 \includegraphics[width=75mm]{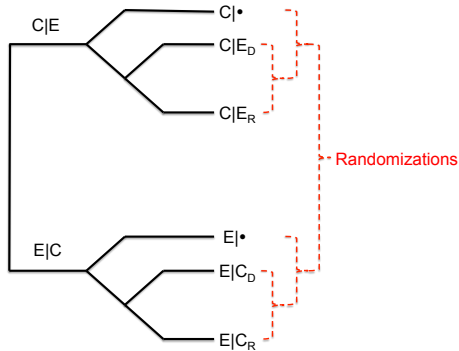}
\end{wrapfigure}

Building upon Smith et al.'s survey, we currently (April 2011) have in the field a pilot instrument designed to measure attitudes toward unified and divided government.  In fashioning the protocols for the study, our central goal was to overcome the obstacles described above.  To that end, our revised instrument asks a ``symmetrical'' set of questions about divided and unified control of the executive and legislative branches of government.  As illustrated in Figure \ref{Qualtrix}, we also randomize question order in blocks: first by institution (executive or legislative), second by conditioning, and finally (within conditional blocks) by identity of partisan control.  In addition, we also include the two-way questions regarding affinity for divided government in general; these questions are also randomized both vis-a-vis their order relative to the items in Figure \ref{Qualtrix} and with respect to each other.  Finally -- building on the early, influential work of Weisberg (1980) -- we measure respondents' party identification through a series of self-identification questions designed so as not to foreclose on the prospect that a subject can maintain an affinity for both parties at the same time.  In each instance, these changes to existing (and ``industry-standard'') protocols will enable us to begin to evaluate quantum-consistent theories of voter attitudes on an equal footing with their classical counterparts.

Beyond overcoming the limitations of existing studies, our broader goals in this endeavor are to create a set of instruments by which we and other researchers can begin to evaluate macro-level quantum and ``pseudo-classical'' processes in sociopolitical data.  In a field (politics) in which the amount and quality of information varies widely, ambivalence of attitudes are widespread, and citizens often exhibit patterns of behavior that fail to correspond to classical models of human action, the intuitions of quantum cognition and action hold particular promise.

\section*{References}

\noindent \bibitem Aerts, D., Aerts, S., Broeckaert, J., and Gabora, L. 2000. ``The Violation of Bell Inequalities in the Macroworld.'' \emph{Foundations of Physics} 30:1378-1414.

\bibitem Aerts, D., and Gabora, L. 2005. ``A Theory of Concepts and Their Combinations II: A Hilbert Space Representation.'' \emph{Kybernetes} 34:176-205. 

\bibitem Born, M. 1926. ``Zur Quantenmechanik der Sto\ss vorg\"ange.'' \emph{Zeitschrift f\"ur Physik} 37: 863-867. 

\bibitem Bruza, P.\ D., Kitto, K., Ramm, B., Sitbon, L., Song, D., and Blomberg, S.  2011.  ``Quantum-like Non-Separability of Concept Combinations, Emergent Associates and Abduction.''  \emph{Logic Journal of the IGPL} forthcoming.
 
\bibitem Bruza, Peter, Azhar Iqbal, and Kirsty Kitto.  2010.  ``The Role of Non-Factorizability in Determining `Pseudo-Classical' Non-Separability.'' \emph{Quantum Informantics for Cognitive, Social, and Semantic Processes: Papers from the AAAI Fall Symposium (QI-2010)}, 26-31. 

\bibitem Busemeyer, J., Pothos, E., Franco, R., and Trueblood, J. 2011. ``A Quantum Theoretical Explanation for Probability Judgment `Errors'.''  \emph{Psychological Review} forthcoming.

\bibitem Campbell, Angus, Philip E.\ Converse, Warren Miller, and Donald Stokes.  1960.  \emph{The American Voter}. New York: Wiley.

\bibitem Carsey, T., and Layman, G.  2004.  ``Policy Balancing and Preferences for Party Control of Government.'' \emph{Political Research Quarterly} 57:541-550.

\bibitem Converse, Philip E.  1964.  ``The Nature of Belief Systems in Mass Publics.''  In David Apter, Ed \emph{Ideology and Discontent}.  New York: Free Press.

\bibitem Davis, O., and Hinich, M. 1968. ``On the Power and Importance of the Mean Preference in a Mathematical Model of Democratic Choice.'' \emph{Public Choice} 5:59-72. 

\bibitem Downs, Anthony.  1957.  \emph{An Economic Theory of Democracy}.  New York: Harper and Row.

\bibitem Enelow, James M., and Melvin J.\ Hinich.  1984.  \emph{The Spatial Theory of Voting: An Introduction}. New York: Cambridge University Press.

\bibitem Fiorina, Morris P.  1996. \emph{Divided Government}, 2nd Ed. New York: MacMillan. 

\bibitem Gabora, L., and D.\ Aerts. 2009. ``A Model of the Emergence and Evolution of Integrated Worldviews.'' \emph{Journal of Mathematical Psychology} 53:434-451.

\bibitem Gorman, W. 1968. ``The Structure of Utility Functions.'' \emph{Review of Economic Studies} 32:369-390. 


\bibitem Hinich, Melvin J., and Michael C.\ Munger.  1997.  \emph{Analytical Politics}.  New York: Cambridge University Press.

\bibitem Lacy, Dean.  2001a.  ``A Theory of Nonseparable Preferences in Survey Responses.'' \emph{American Journal of Political Science} 45:239-258.

\bibitem Lacy, Dean.  2001b. ``Nonseparable Preferences, Measurement Error, and Unstable Survey Responses.'' \emph{Political Analysis} 9:95-115.

\bibitem Lacy, Dean, and Emerson M.\ S.\ Niou.  1998.  ``Elections in Double-Member Districts with Nonseparable Preferences.''  \emph{Journal of Theoretical Politics} 10:89-110.

\bibitem Petrocik, J., and Doherty, J. 1996. ``The Road to Divided Government: Paved without Intention.'' In Peter F.\ Galderisi, Roberta Q.\ Herzberg, and Peter McNamara, eds., \emph{Divided Government: Change, Uncertainty, and the Constitutional Order}. Lanham, MD: Rowman \& Littlefield.

\bibitem Schwartz, T. 1977. ``Collective Choice, Separation of Issues and Vote Trading.'' \emph{American Political Science Review} 71:999-1010.

\bibitem Smith, Charles E.\ Jr., Robert D.\ Brown, John M.\ Bruce, and L.\ Marvin Overby.  1999.  ``Party Balancing and Voting for Congress in the 1996 National Election.'' \emph{American Journal of Political Science} 43:737-764.

\bibitem Trueblood, J., and Busemeyer, J. 2011.  ``A Quantum Probability Explanation for Order Effects on Inference.''  \emph{Cognitive Science}: forthcoming.

\bibitem Weisberg, Herbert.  1980.  ``A Multidimensional Conception of Party Identification.''  \emph{Political Behavior} 2:33-60.

\bibitem Zaller, John, and Stanley Feldman.  1992.  ``A Simple Theory of the Survey Response: Answering Questions versus Revealing Preferences.''  \emph{American Journal of Political Science} 36:579-616.

\end{document}